\def \be {\begin{equation}}
\def \eq {\end{equation}}
\def \bee {\begin{eqnarray}}
\def \eqq {\end{eqnarray}}
\def \nn {\nonumber}
\def \bea {\begin{array}{c}}
\def \eqa {\end{array}}
\def \y {$\;$}

\def \la {\langle}
\def \ra {\rangle}
\def \R {{\bf R}}
\def \C {{\bf C}}

\def \del {\partial}
\def \dels {\partial\kern-.5em / \kern.5em}
\def \As {{A\kern-.5em / \kern.5em}}
\def \Ds {D\kern-.7em / \kern.5em}
\def \Psib {\bar{\Psi}}

\def \cA {{\cal A}}
\def \cH {{\cal H}}
\def \pih {\hat{\pi}}
\def \Dt {\tilde{D}}

\def \a {\alpha}

\def \g {\gamma}
\def \d {\delta}

\def \m {\mu}
\def \n {\nu}
\def \k {\kappa}

\def \L {\Lambda}

\def \r {\rho}
\def \o {\omega}
\def \O {\Omega}

\def \cO {{\cal O}}
\def \cU {{\cal U}}

\documentstyle[12pt,amssymbols]{article}

\begin{document}
\begin{titlepage}
\today          \hfill 
\begin{center}
\hfill    UU-HEP/96-07\\
\hfill    NI9018\\

\vskip .5in

{\large \bf Noncommutative Geometry and D-Branes }

\vskip .5in
Pei-Ming Ho and Yong-Shi Wu \\
\vskip .3in
{\em Department of Physics,
University of Utah \\
Salt Lake City, Utah 84112}
\end{center}

\vskip .5in

\begin{abstract}

We apply noncommutative geometry to a system of  
N parallel D-branes, which is interpreted as a quantum 
space. The Dirac operator defining the quantum 
differential calculus is identified to be the 
zero-momentum mode of the supercharge for strings 
connecting D-branes. As a result of the calculus, 
Connes' Yang-Mills action functional on the quantum 
space reproduces the dimensionally reduced U(N) 
super Yang-Mills action as the low energy effective 
action for D-brane dynamics. Several features that may
look {\it ad hoc} in a noncommutative geometric 
construction are shown to have very natural physical 
or geometric origin in the D-brane picture in 
superstring theory. 

\end{abstract}
\end{titlepage}

\newpage
\renewcommand{\thepage}{\arabic{page}}
\setcounter{page}{1}

\section{Introduction}
D-branes \cite{DLP} are extended dynamical
objects in string theory, on which string endpoints can live 
(having the Dirichlet boundary condition). In recent 
developments, recognition \cite{Pol,PCJ} of these 
nonperturbative degrees of freedom has played a central 
role in understanding  string-dualities, M-theory unification,
and small distance structure of space (or space-time) 
\cite{Reviews,DKPS}. One remarkable feature of D-branes is that 
when there are $N$ parallel D-branes, their coordinates are lifted 
to $N \times N$ matrices \cite{Wit}, and their low-energy dynamics 
is described by dimensional reduction of ten-dimensional 
$U(N)$ super Yang-Mills gauge theory. This reminds us 
of noncommutative geometry \cite{Con}, in which coordinates
as local functions are allowed to be noncommuting. Indeed, 
there are striking similarities between the D-brane dynamics 
and the non-commutative geometric construction of the 
standard model \cite{CL}: the parallel D-branes versus 
the multi-sheet space-time, the inter-brane connections 
versus the Higgs fields, and so on. Moreover, 
noncommutative geometric features also appear 
in a recently conjectured light-cone formulation for 
eleven-dimensional M-theory \cite{BFSS}.  We feel,
as a warm-up for exploring the possible uses of 
noncommutative geometry in string theory including 
M-theory, it is instructive to first examine more 
closely the connection between noncommutative 
geometry and D-brane dynamics. 

In string theory, one is used to start from bosonic degrees 
of freedom. For a D$p$-brane in bosonic string theory,
at low energies the dynamics is described by
two kinds of fields living on the brane.
Let us denote the coordinates on the D-brane by
$y^i$, where $i=0,1,2,\cdots,p$. There is the 
$U(1)$ gauge field $A^i(y)$, $i=0,1,\cdots,p$,
coupled to the motion of string endpoints 
in the tangential directions on the brane.
There is also the Higgs field $\Phi^a(y)$,
$a=p+1,p+2,\cdots, 25$, corresponding to 
vibrations of the D-brane (or the motion of
string endpoints) in the normal directions.
The effective action of a D-brane is the 
Dirac-Born-Infeld action \cite{Lei}, whose leading term
in the gradient expansion is the usual Maxwell action. 
In superstring theory, the content of the fields 
living on a D-brane is enlarged  to include $\Psi$, 
the fermionic super-partners of $A^i$ and $\Phi^a$.

When there are $N$ parallel D-branes (with 
microscopic separations), all fields $A^i$, $\Phi^a$ 
and $\Psi$ become anti-Hermitian $N\times N$ 
matrices \cite{Wit}, describing the effects of short open 
strings ending on different D-branes. At low energies
and to the leading order in the gradient expansion,
the effective action for such a system in superstring 
theory should be the dimensionally-reduced
$U(N)$ super Yang-Mills action \cite{Wit}
\be \label{SUSY-YM}
S= \frac{1}{g}\int d^{p+1}y\;\;
Tr(-\frac{1}{4}F_{\m\n}F^{\m\n}+\frac{i}{2}\Psib\Ds\Psi),
\eq
where $\m,\n=0,1,\cdots,9$, and $\Psi$ is a Majorana-Weyl 
spinor in 10 dimensions.  Both $A_{\m}$ and $\Psi$ are in
the adjoint representation of $U(N)$.
We will use $ i,j,k, \cdots $ for indices of values $0,1,\cdots,p$,
and $a,b,c, \cdots $ for indices of values $p+1,p+2,\cdots,9$.
In this convention, $F_{\m\n}$ splits into
\bee
&F_{ij}=\del_i A_j-\del_j A_i+[A_i,A_j],\\
&F_{ia}=\del_i \Phi_a+[A_i,\Phi_a]\equiv \nabla_i \Phi_a,\\
&F_{ab}=[\Phi_a,\Phi_b],
\eqq
due to dimensional reduction from 10 to $p+1$.
Similarly,
\be
\Ds\Psi=\g^i\nabla_i\Psi+\g^a[\Phi_a,\Psi].
\eq
Explicitly, the low energy effective action for $N$ D-branes is
\bee \label{eff-act}
S&=&\frac{1}{g}\int d^{p+1}y\;\;
Tr(-\frac{1}{4}F_{ij}F^{ij}
-\frac{1}{2}\nabla_i\Phi_a\nabla^i\Phi^a
-\frac{1}{4}[\Phi_a,\Phi_b][\Phi^a,\Phi^b] \nn \\
&&+\frac{i}{2}\Psib(\g^i\nabla_i\Psi+\g^a[\Phi_a,\Psi])).
\eqq

On the other side, Connes' noncommutative geometry 
\cite{Con} generalizes differential calculus and geometry
to spaces, called ``quantum spaces'', on which the algebra 
of functions (including coordinates) is noncommutative. 
In this generalization, ordinary smooth manifolds 
may allow new noncommutative differential calculi.
Noncommutative geometric ideas have been used to
reformulate the action for the standard model  \cite{CL} 
and the $SU(5)$ grand unified theory \cite{CFF1}. 
This is done by starting from a certain noncommutative 
algebra acting on the fermion fields and then introducing 
appropriate Dirac operator to formulate Connes' 
action functional  (including both Yang-Mills and fermion parts) 
on a multi-sheet space-time, with the inter-sheet distances 
directly related to the vacuum expectation values of
the Higgs fields.  In ref. \cite{Cha}, some supersymmetric 
Yang-Mills actions are reformulated as Connes' action 
functional on certiain quantum spaces. In this note, 
we will show that the D-brane action (\ref{eff-act}) can be
rewritten as Yang-Mills-Connes action functional on a 
quantum space representing D-branes.

In Sec.\y\ref{QC-YM} we will first review basics
of quantum differential calculus and Yang-Mills gauge 
theory on a quantum space, and then in Sec.\y\ref{QC-D}
we will consider a certain class of quantum spaces in detail,
which is used in this paper to model a system of $N$ 
D-branes. Subsequently we deduce the Dirac operator 
that defines the desired quantum calculus from the string 
supercharge in Sec.\y\ref{D-S} and find in Sec.\y\ref{D-Q} that 
the corresponding Yang-Mills-Connes functional for this 
calculus is equivalent to the action (\ref{eff-act}). In Sec.\y\ref{D-T} 
we comment on the relation of T-duality with the choice of 
the Dirac operator. In the concluding section, we summarize
in retrospect several features, which look {\it ad hoc} in
a generic noncommutative geometric construction but become 
very natural when put in the context of D-branes in string theory.

\section{Yang-Mills-Connes Functional}
\label{QC-YM}

A quantum space is described by a $*$-algebra\footnote{
An algebra with a map
$*$: $a\rightarrow a^*\in\cA$ which is an anti-involution, i.e.,
$(a^*)^*=a$ and $(ab)^*=b^* a^*$ for $a,b\in\cA$,
is called a $*$-algebra.}
of functions $\cA$ on the quantum space.
A differential calculus on a quantum space
is an extension of $\cA$ to a graded $*$-algebra
$\O^*(\cA)=\oplus_{n=0}\O^{(n)}(\cA)$,
where $\O^{(0)}(\cA)=\cA$ and
$\O^{(n)}(\cA)$ are right $\cA$-modules.\footnote{
That is, if $\o\in\O^{(n)}(\cA)$ then
$\o a\in\O^{(n)}(\cA)$ for all $a\in\cA$.}
An element in $\O^{(n)}(\cA)$ is called
a differential form of degree $n$ or an $n$-form.
The differential algebra $\O^*(\cA)$ also needs to be
equipped with the exterior derivative $d$.
The exterior derivative is a map
$d:\O^{(n)}(\cA)\rightarrow\O^{(n+1)}(\cA)$
satisfying the graded Leibniz rule
$d(\o_1\o_2)=(d\o_1)\o_2+(-1)^{n_1}\o_1(d\o_2)$
for $\o_1\in\O^{(n_1)}(\cA)$ and $\o_2\in\O^*(\cA)$
and the nilpotency condition $d^2=0$.
Typically an element in $\O^{(n)}(\cA)$ can be written as 
$\xi^{\m_1}\cdots \xi^{\m_n}a_{\m_1\cdots\m_n}$
for some $a_{\m_1\cdots\m_n}\in\cA$,
where $\{\xi^{\m}\}$ is a basis of one-forms in $\O^{(1)}(\cA)$.

In Connes' formulation of noncommutative geometry \cite{Con},
all information about a quantum space
is encoded in the spectral triple $(\cA,D,\cH)$,
where $D$ is an anti-self-adjoint operator 
(called the Dirac operator) acting on $\cH$,
which is a Hilbert space with a $*$-representation 
$\pi$ of $\cA$, namely, the $*$-anti-involution 
is realized as the Hermitian conjugation: 
$\pi(a^*)=\pi(a)^{\dagger}$, $\forall a\in\cA$, 
where the Hermitian conjugation is denoted by 
$\dagger$. Using $D$ one can define a noncommutative 
differential calculus $\O^{*}(\cA)$
and extend $\pi$ to a representation $\pih$
of $\O^*(\cA)$ \cite{CFF0}. This procedure is 
shown by an example in Sec.\y\ref{QC-D}.

Other essential ingredients of a quantum space
are the inner product and integration over $\O^*(\cA)$.
Let $\la\cO\ra_{\cH}$ denote the regularized average
of an operator $\cO$ on the Hilbert space $\cH$, e.g.
\be \label{inner}
\la\cO\ra_{\cH}=\lim_{\Lambda\rightarrow\infty}
\frac{Tr_{\cH}(\cO\exp\{-|D|^2/\L^2\})}
{Tr_{\cH}(\exp\{-|D|^2/\L^2\})},
\eq
where $\Lambda$ is the cutoff for the spectrum of the Dirac operator.
The inner product on $\O^{(n)}(\cA)$ is defined by \cite{CFF0}
\footnote{
The inner product and integration can also be defined 
in terms of the Dixmier trace \cite{Con}; what we use 
here is more familiar to physicists. 
}
\be
\la\o_1|\o_2\ra_{\O}=\la\pih(\o_1)^{\dagger}\pih(\o_2)\ra_{\cH}
\eq
for $\o_1,\o_2\in\O^{(n)}(\cA)$.
The integration of $\o\in\O^*(\cA)$ is defined by \cite{CFF0}
\be
\int\o=\la\pih(\o)\ra_{\cH},
\eq
where $\pih$ is a representation of $\O^*(\cA)$ on $\cH$.

The distance between two states
in the Hilbert space is defined by \cite{Con}
\be \label{dist}
dist(\psi_1,\psi_2)
=sup\left\{ \left|\la\psi_1|\pih(a)\psi_1\ra-\la\psi_2|\pih(a)\psi_2\ra\right|:
a\in\cA, \left\| |\pih(da)|^2 \right\|\leq 1\right\},
\eq
where $\|\cdot\|$ is the operator norm and
$\la\cdot|\cdot\ra$ is the inner product on the Hilbert space.

A gauge field theory fits into this framework easily.
The group of unitary elements
$\cU=\{u: uu^* =u^* u=1, u\in\cA\}$
in $\cA$
acts on $\cH$ as a group of transformations,
which is identified with the gauge group.
For example, if $\cA$ is the algebra of $N\times N$
matrices of complex functions on a manifold,
the group $\cU$
is the gauge group $U(N)$ on the manifold.
The gauge field is a one-form
$A=\xi^{\m}A_{\m}\in\O^{(1)}(\cA)$
for some $A_{\m}\in\cA$.
It transforms under $u\in\cU$ as
$A\rightarrow A^u=uAu^*+u(du^*)$,
which implies that
$\pih(A)\rightarrow \pih(A^u)=U\pih(A)U^{\dagger}+U[D,U^{\dagger}]$,
where $U=\pih(u)$ (and $U^{\dagger}=\pih(u^*)$).
The modified Dirac operator
$\Dt=D+\pih(A)$ is covariant.
The gauge-covariant field strength
is defined as usual by $F=dA+A^2$.
The only new ingredient so far
in this straightforward generalization is
the quantum differential calculus
behind each expression above.
The Yang-Mills-Connes action functional
defined by \cite{Con}
\be \label{YM-M}
S=\la F|F \ra_{\O} + \la\Psi|\Dt\Psi\ra,
\eq
where $\la\cdot|\cdot\ra$ is the inner product on $\cH$,
is another natural but non-trivial ingredient
of the noncommutative generalization.

For a non-Abelian gauge field $A=dx^{\m}A_{\m}$
on a classical manifold,
the distance defined
by (\ref{dist}) with $\Dt=\dels+\As$
between two vectors in the fibers
located at two points on the manifold is the length
of the shortest path on the manifold which connects
the two vectors by parallel transport \cite{Con1}.

\section{A Class of Quantum Calculi}
\label{QC-D}

A basic idea in applying noncommutative geometry to
field theory is that the geometry of the relevant quantum 
space is determined by matter or, more precisely, by 
fermion fields. Thus one is led to consider a special class 
of quantum calculi, where the $*$-algebra $\cA$ is 
a noncommutative algebra acting on fermion fields, and the 
Dirac operator acting on the fermions is of the form
\be \label{gD}
D=\g^{\m}\otimes D_{\mu},
\eq
where the $\g^{\m}$'s are usual $\g$-matrices.
The Hilbert space $\cH$ is one for fermions, 
of the form $\cH=S\otimes\cH_0$, where $S$ is a 
representation of the Clifford algebra (e.g. 
$S=\C^{32}$ for a Dirac spinor in 10 dimensions), 
and $\cH_0$ is the Hilbert space in which the 
algebra $\cA$ acts with a representation $\pi_0$.
The representation of $a\in\cA$ on $\cH$ is
$\pi(a)=1\otimes\pi_0(a)$. From now on we will 
suppress the symbol of tensor product $\otimes$.

In the universal differential calculus $\O^*\cA$ \cite{Con}, 
a differential one-form $\r\in\O^{(1)}\cA$
is a formal expression $\r=\sum_{\a}a_{\a}db_{\a}$,
where $a_{\a}$, $b_{\a}$ are elements in $\cA$.
To simplify the notation,
we will omit the index $\a$ in the following.

With the help of the Dirac operator,
the representation $\pi$ of $\cA$ on $\cH$
is extended to $\O^*\cA$ \cite{Con}
by defining, for $\r\in\O^{(1)}\cA$,
\footnote{We will simply write $a$ to stand for $\pi(a)$
for $a\in\cA$ in the following.}
\be
\pi(\r)=\sum a[D,b].
\eq
By (\ref{gD}), it is
$
\pi(\r)=\g^{\m}\sum a[D_{\m},b]=\g^{\m}\r_{\m}.
$
The representation of a two-form $\o=\sum adbdc$ is thus
\be
\pi(\o)=\sum a[D,b][D,c],
\eq
and similarly for forms of higher degrees.

In particular, the representation of $d\r=\sum dadb$ is
\bee \label{dr}
\pi(d\r)
&=&\g^{\m\n}
([D_{\m},\r_{\n}]-\frac{1}{2}\sum a[[D_{\m},D_{\n}],b]) \nn \\
&&+g^{\m\n}([D_{\m},\r_{\n}]-\sum a[D_{\m},[D_{\n},b]]),
\eqq
where $\g^{\m\n}=\frac{1}{2}[\g^{\m},\g^{\n}]$
and $g^{\m\n}=\frac{1}{2}\{\g^{\m},\g^{\n}\}$
is the metric.

The general differential calculus $\O^*(\cA)$ \cite{Con}
is defined by the quotient
\be
\O^*(\cA)=\O^*\cA/J,
\eq
where $J=ker\pi + d(ker\pi)$.
This means that two differential forms $\o_1$ and $\o_2$
of the same degree will be considered the same
if $(\o_1-\o_2)\in J$.

To find $\O^{(2)}(\cA)$, the differential calculus of degree two,
we consider a one-form $\r\in\ker\pi$.
According to (\ref{dr}), if the zero-curvature condition 
for the unperturbed Dirac operator
\be \label{DD}
[D_{\m},D_{\n}]=0
\eq
is satisfied,
then
\be \label{dr1}
\pi(d\r)=-\sum a[g^{\m\n}D_{\m}D_{\n},b]\in\pi(\cA).
\eq
For a non-Abelian gauge theory on a classical manifold,
$D_{\m}$ can be $\del_{\m}$ plus a pure gauge to satisfy (\ref{DD}).

Denote the degree-two component of $J$ by $J^{(2)}$, then
$\pi(J^{(2)})$
is composed of all elements
(\ref{dr1}) for all $\r=\sum adb\in ker\pi$.
We will focus on the cases for which
$D_{\m}$ satisfies (\ref{DD})
and
\be \label{JA}
\pi(J^{(2)})=\pi(\cA).
\eq

The representation $\pi$ defined above is, in general,
not a good representation of $\O^*(\cA)$
because one and the same differential form may admit many
equivalent expressions of the form $\sum adbdc\cdots$,
so that the representation is not unique.

A good representation is given by
$\pih=P_J\circ\pi$ \cite{CFF0} where $P_J$
is the projection
perpendicular to $\pi(J)$.
By (\ref{dr}) and (\ref{JA}),
it follows that for a two-form $\o$,
\be
\pih(\o)=\g^{\m\n}\o_{\m\n}
\eq
for some $\o_{\m\n}\in\cA$.
In particular, by (\ref{DD}),
\be \label{dr2}
\pih(d\r)=
\frac{1}{2}\g^{\m\n}([D_{\m},\r_{\n}]-[D_{\n},\r_{\m}]).
\eq

It can be shown that the conditions (\ref{DD}) and (\ref{JA})
also imply that for a three-form $\o$,
\be
\pih(\o)=\g^{\m\n\k}\o_{\m\n\k}
\eq
for some $\o_{\m\n\k}\in\cA$,
and similarly for higher degrees.

Let $\xi^{\m}$ denote the basis of one-forms which
is represented by $\g$-matrices: $\pi(\xi^{\m})=\g^{\m}$.
Then it follows that the calculus $\O^*(\cA)$
is generated by elements in $\cA$ and one-forms $\xi^{\m}$,
where the $\xi^{\m}$'s anticommute with each other
and commute with elements in $\cA$
as in the classical case. The only possible source of 
noncommutativity is $\cA$.

\section{Dirac Operator and Supercharge}
\label{D-S}

The low-energy dynamics of $N$ parallel D-branes
is described by a field theory on the D-brane world 
volume. So we may try to reformulate the D-brane
action (\ref{eff-act}) in terms of the quantum 
calculus discussed in last section. The key is to find an
appropriate Dirac operator for the fermion fields, 
which are massless spinor states of strings
connecting the D-branes. The Dirac operator
should be motivated from string theory: Indeed
we find a natural candidate to be the supercharge 
operator for strings connecting D-branes, truncated
in the subspace of massless spinor states. 
 
It was shown by Witten \cite{Wit82} that the quantized 
zero-momentum modes in a supersymmetric 
non-linear $\sigma$-model in $1+1$ dimensions 
can be identified with the de Rham complex
of the target space. The bosonic fields $X^{\m}$
are the coordinates on the target space.
The fermionic fields $\psi^{\m}$ are Majorana spinors
on the world sheet,
which splits into two Majorana-Weyl spinors
$\psi^{\m}_+, \psi^{\m}_-$ in $1+1$ dimensions.
By canonical quantization,
$\psi^{\m}_+$ and $\psi^{\m}_-$ satisfy
two anticommuting sets of Clifford algebra:
$\{\psi^{\m}_A,\psi^{\n}_B\}=g^{\m\n}\d_{AB}$,
where $A, B=+, -$ and
$g^{\m\n}$ is the metric of the target space.
The supercharge $Q$ on the world sheet for zero-momentum modes
is also a Majorana spinor
and has two Weyl components $Q_+$ and $Q_-$:
\be
Q_{\pm}=\psi^{\m}_{\pm}P_{\m},
\eq
where the momentum $P^{\m}$ acts on functions of $X^{\m}$
as a derivative.
Let $Q=\frac{1}{2}(Q_+ +iQ_-)$
and $Q^*=\frac{1}{2}(Q_+ -iQ_-)$.
It is remarkable that the supercharges $Q$ and $Q^*$
realize the exterior derivative $d$
and its adjoint $d^*$ \cite{Wit82}.
$\psi^{\m}=\psi^{\m}_+ +i\psi^{\m}_-$ and
$\psi^{\m *}=\psi^{\m}_+ -i\psi^{\m}_-$
correspond to differential one-forms
and inner derivatives, respectively.
Hermitian conjugation realizes
Poincar\'{e} duality.

All these are also true for closed strings.
For an open string with Neumann boundary conditions,
however, certain modification is necessary, because the 
zero-momentum modes of the right-moving and
left-moving sectors of $\psi^{\m}$ are identified.
Then there is only one set of Clifford algebra,
and the supercharge of zero-momentum modes
has only one independent component
\be \label{Q-2}
Q_0=\psi^{\m}_0 P_{\m}.
\eq
After canonical quantization,
$\psi^{\m}_0$'s become $\g$-matrices acting
on massless spinor states,
which are (after GSO projection)
Majorana-Weyl spinors
in the supermultiplet of a Yang-Mills theory
in 10 dimensions \cite{GSW}.
Being the Dirac operator for these spinors,
the supercharge $Q_0$
realizes the exterior derivative
in the target space in the sense of Connes.

Later it has also been argued by Witten 
\cite{Wit85} that the generalized Dirac 
operator in the full superstring theory
(including nonzero-momentum sector),
the so-called Dirac-Ramond operator,
is the supercharge on the world sheet
for the following three reasons:
(1) Its zero-momentum limit is the usual Dirac operator.
(2) It annihilates physical states.
(3) It anticommutes with an analogue of
the chirality operator: $(-1)^F$.

Motivated by these observations, we try to deduce in 
the following the Dirac operator for the quantum 
space representing D-branes from the supercharge 
on the world sheet of strings ending on D-branes.

It is well known \cite{DLP} that for open strings ending 
on a D$p$-brane, we have for $X^a$, $a=p+1,\cdots,9$,
Dirichlet boundary conditions, which originate from
$T$-duality for Neumann boundary conditions.
Since $T$-duality simply reverses the relative sign on
the left-moving and right-moving modes
for both $X^{\m}$ and $\psi^{\m}$,
the appropriate boundary conditions for $\psi^{\m}$
on a string with Dirichlet boundary conditions
are still \cite{PCJ}
\be
\psi^{\m}_+(0,\tau)=\psi^{\m}_-(0,\tau),\quad
\psi^{\m}_+(\pi,\tau)=\pm\psi^{\m}_-(\pi,\tau),
\eq
(except changes in sign for dualized directions)
of the same type as for open strings with Neumann boundary 
conditions. As mentioned above, the supercharge 
(\ref{Q-2}) acts on the states of massless spinors.
Upon identifying these massless string states 
with the field $\Psi$ in the super Yang-Mills 
theory on a 9-brane, the string supercharge reduces
(or truncates) to the Dirac operator in 10 dimensional 
spacetime. For a D-brane of lower dimensions, the 
momentum operators in directions normal to the brane 
vanish due to the Dirichlet boundary conditions,
hence the supercharge becomes a Dirac operator
on the $(p+1)$ dimensional world volume of the 
D$p$-brane. However, to fully describe the dynamics 
of massless fields on a D-brane, one has to include the 
effects of the tadpole diagram for closed strings
created from the D-brane. Although it was mentioned 
before that the supercharge has two components $Q,Q^*$
on a closed string, the boundary conditions on the brane
identify $\psi^{\m}$ and $\psi^{\m*}$ up to a 
sign \cite{Pol} (after all, a closed string tadpole 
diagram can be viewed as an open string disk diagram),
so half of the supersymmetry is broken.
Hence the D-brane is a BPS state \cite{Pol} and
there is only one independent component
of the supercharge on a closed string created from 
a D-brane. The momentum of the closed string
is shifted by the gauge field $\phi^a$ normal
to the brane, which originates from
a pure gauge transformation $\L^a=\phi^a x^a$ 
in the dual picture. Including contributions 
from both open and closed strings, the Dirac operator 
obtained by truncating the supercharge on 
strings ending on a D-brane is
\be\label{old-Dirac}
D=\g^i\partial_i+\g^a\phi_a.
\eq

\section{D-branes as Quantum Space}
\label{D-Q}

Now we are able to formulate precisely how to interpret  
the system of $N$ parallel D-branes (with microscopic 
separations) as a quantum space. We take a $p+1$
dimensional  coordinate system on one of the branes
as the world volume coordinates, and treat the structure 
arising from the strings connecting D-branes 
as the ``internal'' structure that defines a quantum space.
 (Closed string tubes connecting two branes can be 
viewed as loops of open strings ending on different branes.) 
Each D-brane has a label $r$ ($r=1,2,\cdots, N$), and an 
open (oriented) string from D-brane $r$ to D-brane $s$,
and the states on such string may be labeled by an ordered
pair, $(r,s)$, of indices. (These indices are also called 
Chan-Paton labels, since such an open string is dual to an 
open string with usual Chan-Paton labels $(r,s)$ \cite{PCJ}.) 
In particular, the massless spinor states of the open string 
connecting D-brane $r$ and D-brane $s$ result in 
fermion fields living on the D-brane world volume, 
which therefore also carry the Chan-Paton labels $(r,s)$.
Thus the fermion field $\Psi$ is an $N\times N$ matrix 
with entries being Majorana-Weyl spinors in 10 dimensions, 
which are naturally anti-Hermitian, belonging to the adjoint 
representation of $U(N)$: exchanging the labels $r$ and 
$s$ leads to inverting the orientation of the string. 
The Hilbert space on which the Dirac operator acts is 
taken to be the space of $N\times N$ matrices of Dirac 
spinors, which is larger than the configuration space of 
$\Psi$, since the Dirac operator always reverses the chirality.

To define a quantum space representing the D-branes, 
in addition to $\cH$ we need to specify the other two
elements in the spectral triple $(\cA,D,\cH)$. 
Recall that the algebra $\cA$ defines the gauge group
as the group $\cU$ of unitary elements of $\cA$.
Hence we take $\cA$ to be $M_N(\C)\otimes 
L^2(\R^{p+1})$, the algebra of $N\times N$ matrices 
of square-integrable functions on the $p+1$ dimensional 
D-brane world volume,
\footnote{It is also possible to define $\cA$ to be
the algebra of the $U(N)$ gauge group represented
in its adjoint representation on $\cH$.}
so that $\cU$ is the $U(N)$ 
gauge group. The representation $\pi$ of $\cA$ on 
the fermion Hilbert space $\cH$ is simply the matrix 
multiplication. The Dirac operator is chosen to be
a natural generalization of the operator (\ref{old-Dirac})
to the multi-brane case. By introducing a Wilson line
$A^a=\phi_a\equiv diag(\phi^a_1,\cdots,\phi^a_N)$ 
in the T-dual picture, the Dirac operator resulting 
from the (truncated) supercharge operator is found to be
\be \label{old-Dirac2}
D=\g^i\partial_i+\g^a\phi_a,
\eq
where the $\phi_a$'s are $N\times N$ matrices,
automatically satisfying
\be \label{phiphi}
[\phi_a,\phi_b]=0, \quad \forall a,b=p+1,\cdots,9.
\eq
The first term in eq. (\ref{old-Dirac2}) is the classical 
Dirac operator on the $(p+1)$ dimensional D-brane world 
volume. Viewing a $p$-brane as dimensional reduction of
a $9$-brane, one can think of the second term as
the remnant of the dimensionally reduced $(9-p)$ directions
along which the partial derivatives $\del_a$ vanish but the 
pure gauge terms survive. (Recall the statement following 
(\ref{dr1}).)
 

The quantum calculus considered in Sec.\y\ref{QC-D}
is applicable to the present case. Now let us 
show that the Yang-Mills-Connes 
functional (\ref{YM-M}) with the Dirac operator 
(\ref{old-Dirac2}) reproduces the super Yang-Mills 
action (\ref{eff-act}) describing the dynamics of $N$ D-branes.

The Dirac operator (\ref{old-Dirac2}) satisfies (\ref{DD})
because of (\ref{phiphi}), and for generic $\phi^a$ it also 
satisfies (\ref{JA}),
thus according to the 
discussions in Sec.\y\ref{QC-D}, for generic $\phi^a$
the calculus $\O^*(\cA)$ on D-branes is generated by 
$\cA$ and $dx^{\m}$.
\footnote{
Strictly speaking, our notation for $dx^a$ ($a=p+1,\cdots,9$)
is inappropriate
because while $dx^i$ ($i=0,\cdots,p$) is exact, 
$dx^a$ is not exact but a closed one-form $\sum_{\a}a_{\a}db_{\a}$
for $a_{\a}$, $b_{\a}$ being some matrices in $\cA$.
}
The only noncommutativity resides in $\cA$, the algebra 
of matrices $M_N(\C)\otimes L^2(\R^{p+1})$.
The one-forms $dx^{\m}$ are represented by
$\g$-matrices: $\pih(dx^{\m_1}\cdots dx^{\m_k})=\g^{\m_1\cdots\m_k}$
and so they anticommute with each other
and commute with elements in $\cA$.

The gauge field is a one-form
\be
A=dx^i A_i+dx^a A_a,
\eq
where $A_i$ and $A_a$ are required to be anti-Hermitian.
It modifies the Dirac operator to
\be \label{new-Dirac}
\Dt=\g^i\nabla_i+\g^a\Phi_a,
\eq
where $\nabla_i=\del_i+A_i$
and
\be
\Phi_a=\phi_a+A_a.
\eq

The field strength, 
\be
F=\frac{1}{2}dx^{\m}dx^{\n}F_{\m\n}=dA+A^2, 
\eq
is given by
\bee
&F_{ij}=\del_i A_j-\del_j A_i+[A_i,A_j], \\
&F_{ia}=\del_i \Phi_a+[A_i,\Phi_a]\equiv\nabla_i\Phi_a, \\
&F_{ab}=[\Phi_a,\Phi_b]-[\phi_a,\phi_b]=[\Phi_a,\Phi_b],
\eqq
where we have used (\ref{phiphi}).

The first term in the Yang-Mills functional (\ref{YM-M})
involes the trace of the Hilbert space in (\ref{inner}),
which is composed of three kinds of traces.
The first is the trace of $\g$-matrices,
which gives rise to the contraction of the components of
two $F_{\m\n}$. The second is the trace over
square-integrable functions on $\R^{p+1}$, which turns 
into the integration over the world volume of the D$p$-brane. 
The trace of $N\times N$ matrices remains explicit as 
in (\ref{eff-act}). It is then straightforward to see that
by constraining physical states to be anti-Hermitian 
$N\times N$ Majorana-Weyl spinors $\Psi$ after Wick rotation,
the Yang-Mills functional (\ref{YM-M}) for this quantum space is
equivalent to the effective action (\ref{eff-act}) for $N$ D-branes.
Obviously the same formulation can be applied to other cases, 
for example, the $N=2$ super Yang-Mills theory in four dimensions
as dimensionally reduced from $N=1$ super Yang-Mills theory
in six dimensions.

\section{T-duality and Dirac operator}
\label{D-T}

Using T-duality,
we can define another Dirac operator
by taking the T-dual of the supercharge
on an open string with Neumann boundary conditions.
Introducing a Wilson line in the dual picture:
$A^a=diag(\phi^a_1,\cdots,\phi^a_N)$
in some compactified dimensions of radius $R^a$,
the open string with Chan-Paton labels $(r,s)$
is T-dual to an open string stretching between two D-branes
at positions $x^a_r=\phi^a_r$
and $x^a_s=\phi^a_s$
(or a simultaneous translation of them) \cite{PCJ}.
The momentum $p^a$ in the compactified direction
is shifted by the gauge field:
\be
p^a=\frac{n^a}{R^a}+(\phi^a_s-\phi^a_r),
\eq
where $n$ is the quantum number for momentum $p^a$
in the dual picture and becomes the winding number
in a compactified dimension of radius $R'^a=\a'/R^a$.
As we are focusing on low energy modes, we set $n^a=0$.

To describe all string states at the same time,
it is natural to put the $p^a$'s
for all possible string configurations
into an antisymmetric matrix
\be
P^a_{rs}=\phi^a_s-\phi^a_r,
\eq
which is in the Lie algebra of $SO(N)\subset U(N)$.
The Dirac operator as the ``total'' supercharge (\ref{Q-2})
therefore becomes
\be
D_{dual}=\g^i\partial_i+\g^{a}P_a,
\eq
where $i=0,\cdots,p$, and $a=p+1,\cdots,9$.

Consider the case of two D$8$-branes ($N=2$).
The matrix $P^a$ ($a=9$) is
\be
\left(
\begin{array}{cc}
0 & (x^a_1-x^a_2) \\
(x^a_2-x^a_1) & 0
\end{array}
\right).
\eq
When the distance $|x^a_1-x^a_2|$ between two branes
is large, the gauge group is $U(1)^2$.
Hence the algebra of functions $\cA$ is taken to be
diagonal $2\times 2$ matrices for this case.
This is precisely the two-sheet model Connes considered \cite{Con}
and the distance (\ref{dist}) for this case is
$|(x^a_1-x^a_2)|^{-1}$, the inverse of the actual distance.
This is not surprising because we are using
the Dirac operator obtained from the supercharge 
in the dual picture and T-duality inverses the length.
It is interesting to note that if we take the inverse
of every element of $P^a$ to ``correct'' this inversion 
in length: i.e. use 
\be
P'^a_{rs}=(x^a_s-x^a_r)^{-1}
\eq
to replace $P^a$ in the Dirac operator
for $N$ D$8$-branes,
then the new Dirac operator will define the geometry
of $N$-sheets separated by the actual distances
$|(x^a_r-x^a_s)|$.
The algebra $\cA$ in this case is the algebra of
diagonal $N\times N$ matrices,
as appropriate for the $U(1)^N$ gauge symmetry.

\section{Discussions}

In this paper, we have interpreted the system of N parallel 
D-branes as a quantum space in the sense of 
noncommutative geometry. The associated
Yang-Mills-Connes action functional on this 
quantum space is shown to reproduce the dimensionally 
reduced U(N) super Yang-Mills action as the low 
energy effective action for D-brane dynamics. To conclude, 
in this section we note in retrospect that 
several features that would look {\it ad hoc} in 
a noncommutative geometric construction actually 
have very natural physical or geometric interpretation 
in the D-brane picture in string theory. 

First, the source of noncommutativity resides in the 
matrix algebra $\cA$, which arises naturally due to the 
Chan-Paton labels of the fermion fields, which 
in turn originate from the strings ending on 
different D-branes. In other words, parallel 
D-branes provide a physical realization of ``multi-sheet 
space-time'' and a geometric origin for the gauge group
$U(N)$. One may wonder whether our universe could 
really be such a system of D-branes or, equivalently, 
have spacetime of a discrete Kaluza-Klein structure.

Second, the choice of the Dirac operator  
(\ref{old-Dirac2}) is dictated by the D-brane
picture, where the addition of the second term is
due to the fluctuations in the position of the
D-branes. In particular, the commutativity  (\ref{phiphi}) 
that makes the condition (\ref{DD}) satisfied is 
not an ansatz as in usual noncommutative geometric 
reformulation of super Yang-Mills action \cite{Cha};
it is deduced here from T-duality of the D-branes:
the inter-brane separations is dual to a Wilson line for
pure gauge configuration \cite{PCJ,Wit}.

Third, in the Yang-Mills-Connes action functional
(\ref{eff-act}), the $\phi_a$ that is introduced in 
the unperturbed Dirac operator (\ref{old-Dirac2}) 
appears only in the combination $\Phi_a 
= A_a + \phi_a$. In the D-brane picture 
this is a reflection of the fact that 
$\phi_a$ stands for classical inter-brane separation,
while $A_a$ its quantum fluctuations, as is consistent 
with $\phi_a$ being diagonal and constant and with 
the commutativity constraint (\ref{phiphi}). In 
accordance to T-dulaity, in string theory 
it is the total $\Phi_a$ (together with $A_i$) 
that stands for the D-brane ``coordinates'' 
(divided by $\alpha'$, the string tension) 
lifted to a matrix \cite{Wit}. 
We note that such interpretation is not available 
in usual noncommutative geometric construction.

Finally, in general the Yang-Mills-Connes action 
functional is not necessarily supersymmetric. 
However, in the present case, our Yang-Mills-Connes 
action functional (\ref{eff-act}) happens to be 
supersymmetric. This is closely related to the fact
that we start with a very special fermion field content
in a special dimensionality (dimensional reduction of a 
Majorana-Weyl spinor in ten dimensions), which is 
inherited from superstring theory.  

{}From the above discussions, we see that there is
a close relationship and deep internal consistency 
between noncommutative geometry (at least on 
discrete Kaluza-Klein space-time) and D-brane 
dynamics at low energies. An interesting question 
arises: whether or not this close relationship 
of D-brane dynamics with noncommutative geometry 
can be extended to a deeper level? (Either to the 
full D-brane dynamics which should be described by 
a supersymmetric and non-abelian generalization of 
the Dirac-Born-Infeld action, or to superstring 
theory or even M-theory.) This seems to call for a
generalization of noncommutative geometry to 
superstrings or M-theory that incorporates D-branes.

\section{Acknowledgement}

P.M.H. is grateful to Bruno Zumino for encouragements,
Zheng Yin for discussions and the hospitality of
the Isaac Newton Institute for Mathematical Sciences
at the University of Cambridge.
This work is supported in part by
the Rosenbaum fellowship (P.M.H.) and
by U.S. NSF grant PHY-9601277.

\vskip .8cm

{\it Note Added}:
When we are completing the paper, we learn that in a recent
preprint of M. Douglas, hep-th/9610041, a comparison between
the D-brane action and noncommutative geometric construction 
of the standard model action is briefly discussed (without 
much detail).


\baselineskip 22pt

\begin{thebibliography}{10}

\itemsep 0pt

\bibitem{DLP}
J. Dai, R. G. Leigh, J. Polchinski:
``New Connections Between String Theories'',
Mod. Phys. Lett. {\bf A 4}, 2073-2083 (1989).

\bibitem{Pol}         
J. Polchinski:
``Dirichlet-Branes and Ramond-Ramond Charges'',
Phys. Rev. Lett. {\bf 75}, 4724 (1995),
hep-th/9510017.

\bibitem{PCJ}
J. Polchinski, S. Chaudhuri, C. V. Johnson:
``Notes on D-Branes'', NSF-ITP-96-003 (1996),
hep-th/9602052.

\bibitem{Reviews} 
For recent reviews see, e.g. 
J. H. Schwarz, ``Lectures on Superstring and M Theory
Dualities'', hep-th/9607201;
M. J. Duff, ``M-Theory (the Theory Formerly Known 
as Strings)'', hep-th/9608117;
A. Sen, ``Unification of String Dualities'', 
hep-th/9609176.

\bibitem{DKPS}
M. R. Douglas, D. Kabat, P. Pouliot and S. Shenker,
``D-branes and Short Distance in String Theory'',
hep-th/9608024.


\bibitem{Wit}
E. Witten:
``Bound States of Strings and p-Branes'',
Nucl. Phys. {\bf B 460}, 335-350 (1996).

\bibitem{Con}
A. Connes: ``{\it Noncommutative Geometry}'',
Academic Press (1994).

\bibitem{CL}
A. Connes, J. Lott:
``Particle Models and Noncommutative Geometry'',
Nucl. Phys. {\bf B} {\em (Proc. Suppl.)} {\bf 18}, 29-47 (1990).

\bibitem{BFSS}
T. Banks, W. Fischler, S. H. Shenker, L. Susskind:
``M Theory as a Matrix Model: A Conjecture'',
hep-th/9610043.

\bibitem{Lei}
R. G. Leigh:
``Dirac-Born-Infeld Action From Dirichlet $\sigma$-model'',
Mod. Phys. Lett. {\bf A 4}, 2767-2772 (1989).

\bibitem{CFF1}
A. H. Chamseddine, G. Felder, J. Fr\"{o}hlich:
``Grand Unification in Noncommutative Geometry'',
Nucl. Phys. {\bf B 395}, 672-698 (1993).

\bibitem{Cha}
A. H. Chamseddine:
``Connection Between Space-Time Supersymmetry and
Non-Commutative Geometry'',
Phys. Lett. {\bf B 332} 349-357 (1994).

\bibitem{CFF0}
A. H. Chamseddine, G. Felder, J. Fr\"{o}hlich:
``Gravity in Non-Commutative Geometry'',
Comm. Math. Phys. {\bf 155}, 205-217 (1993).

\bibitem{Con1}
A. Connes:
``Gravity Coupled With Matter and the Foundations
of Noncommutative Geometry'',
Comm. Math. Phys. {\bf 155}, 109 (1996).

\bibitem{Wit82}
E. Witten:
``Constraints on Supersymmetry Breaking'',
Nucl. Phys. {\bf B 202}, 253-316 (1982).

\bibitem{GSW}
M. B. Green, J. H. Schwarz, E. Witten:
``Superstring Theory'',
Cambridge Univ. Press (1987).

\bibitem{Wit85}
E. Witten:
``Global Anomalies in String Theory'',
in Symposium on Anomalies, Geometry and Topology,
pp.61-99, W. A. Bardeen, A. R. White (ed.),
World-Scientific (1985).


\end{thebibliography}
\end{document}